\documentclass[reprint,prl]{revtex4-1}
\usepackage{amssymb,amsmath}
\usepackage{graphicx}
\usepackage{dcolumn}
\usepackage{multirow}
\usepackage{color}
\usepackage[english]{babel}
\usepackage[caption=false]{subfig} 

\newcommand{\ket}[1]{\ensuremath{|{#1}\rangle}}

\begin{document}

\def\simlt{\mathrel{\lower .3ex \rlap{$\sim$}\raise .5ex \hbox{$<$}}}

\title{\textbf{\fontfamily{phv}\selectfont 
Fast coherent manipulation of three-electron states in a double quantum dot
}}
\author{Zhan Shi}
\author{C. B. Simmons}
\author{Daniel R. Ward}
\author{J. R. Prance}
\author{Xian Wu}
\author{Teck Seng Koh}
\author{John King Gamble}
\author{D. E. Savage}
\author{M. G. Lagally}
\author{Mark Friesen}
\author{S. N. Coppersmith}
\author{M. A. Eriksson}
\affiliation{University of Wisconsin-Madison, Madison, WI 53706}

\maketitle




\textbf{A fundamental goal in the manipulation of quantum systems is the achievement of many coherent oscillations within the characteristic dephasing time T$_2^*$ \cite{DiVincenzo:2000p1536}.  Most manipulations of electron spins in quantum dots have focused on the construction and control of two-state quantum systems, or qubits, in which each quantum dot is occupied by a single electron
\cite{Petta:2005p2180,Koppens:2006p766,Gaudreau:2011p54,Shulman:2012p202,Maune:2012p344,Medford:2013preprint}.
Here we perform quantum manipulations on a system with
more electrons per quantum dot, in a double dot with
three electrons.
We demonstrate that tailored pulse sequences can be used to
induce coherent rotations between 3-electron quantum states.
Certain pulse sequences yield 
coherent oscillations with a very high figure of merit (the ratio
of coherence time to rotation time) of $>$~100.
The presence of the third electron
enables very fast rotations to all possible states, in contrast to the case when only two
electrons are used, in which some rotations are slow.  The minimum oscillation frequency we observe is $>$~5~GHz.}


Electrons in semiconductor quantum dots are promising candidates for use in quantum computing, because of the potential of this platform to enable coherent quantum control on large numbers of qubits~\cite{Loss:1998p120}.
Much recent progress has led to demonstrations of both spin- and charge-based qubits in both GaAs and Si 
\cite{Hayashi:2003p226804,Petta:2005p2180,Koppens:2006p766,Shulman:2012p202,Morello:2010p687,Maune:2012p344,Petersson:2010p246804,Dovzhenko:2011p161802,Shi:2012preprint,Gaudreau:2011p54,Medford:2013preprint}.
Charge qubits can be manipulated quickly but have relatively short coherence times, while spin
qubits have long coherence times but long manipulation times.
The tendency of the speed of manipulation to be correlated with the rate of decoherence is not
surprising, because
both depend on the coupling of the qubit to external degrees of freedom 
(designed and wanted for manipulation, and extraneous and unwanted for decoherence).
Refs.~\cite{Shi:2012p140503,Koh:2012p250503}, describing a quantum dot hybrid qubit, present theoretical arguments
 that a system with more degrees of freedom, specifically three electrons
in two quantum dots, can overcome this tendency.
Two of the states can form a qubit with spin character that has a long
coherence time.
By accessing a third state via a charge transition, 
fast operations can be performed,
and then the qubit can be converted back into a spin-like qubit with long
coherence time.
Such a strategy requires that one can systematically and accurately
control transitions
between several different quantum states of the same system.

This paper presents experiments that demonstrate the ability to 
tailor transitions between quantum states of
 three electrons in two quantum dots (see Methods for details of device fabrication).
 Four states are important in this work, the ground \ket{0} and first excited state \ket{1} of the dot in the (2,1) charge occupation, and the corresponding ground \ket{2} and first excited state \ket{3} of the dot in the (1,2) charge occupation.
 The qubits are manipulated by pulsing a voltage that changes
 the detuning $\varepsilon$, which is the energy difference between the two dots.
By applying appropriate sequences of applied voltage pulses, oscillations
between different pairs of quantum levels can be induced.
Because oscillations with periods much shorter than the rise times
of the applied pulses can be excited, and because
quantum oscillations between some pairs of the states 
are quite insensitive to the dominant
dephasing mechanism, which is fluctuations in the value of the detuning \cite{Taylor:2007p464},
many (over a hundred) oscillations can be observed within one coherence time.
The consistency of our interpretation of the data in terms of coherent quantum oscillations
between different energy levels is demonstrated by the agreement between the data,
which were all taken with one tuning of the dot, and the
simulations shown, which were all performed with one set of values for the system parameters.

\begin{figure*}
\vspace{-10pt}
\includegraphics[width=\textwidth]{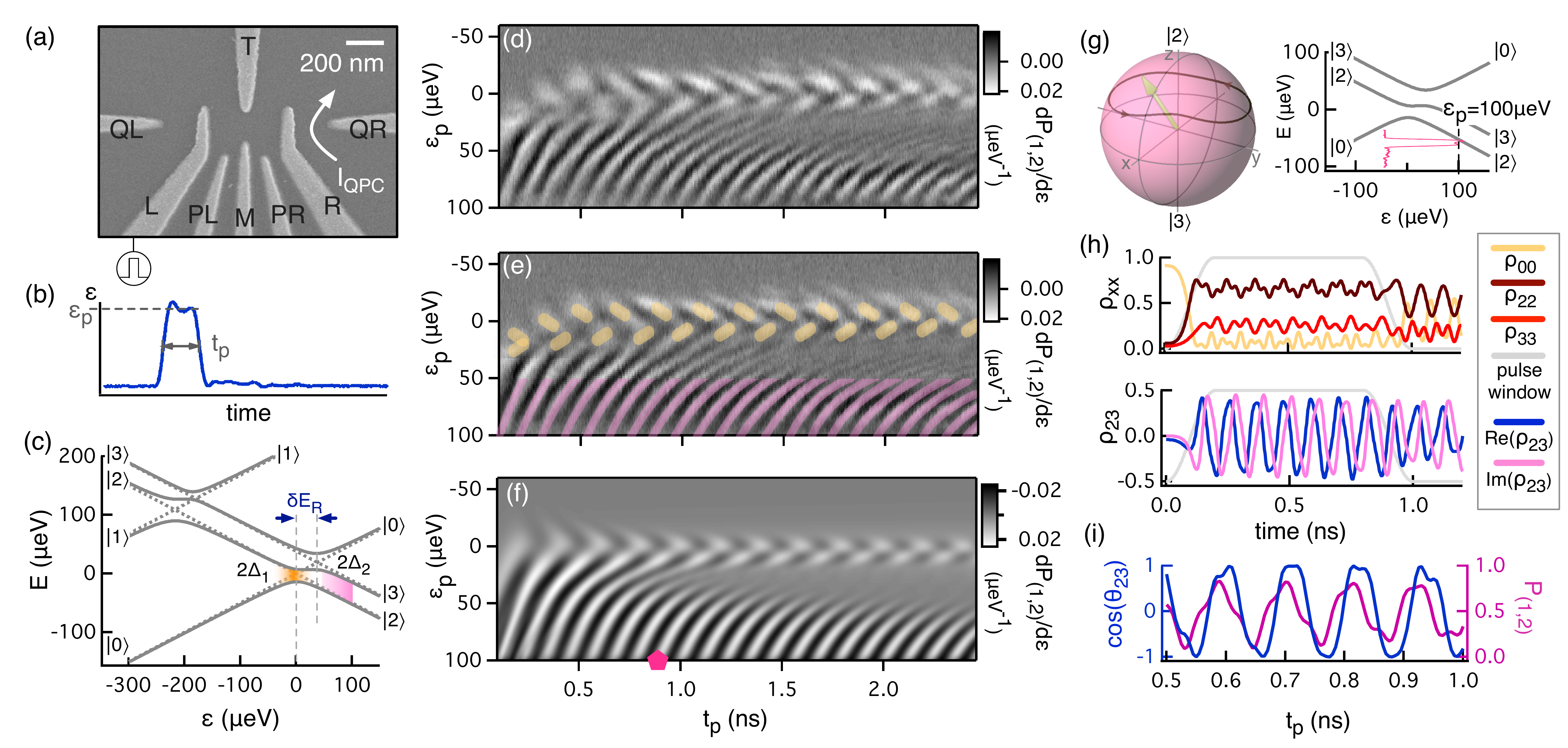}
\caption{Measurement of
quantum oscillations between three quantum states of three electrons in two quantum dots.
(a) Scanning electron micrograph of a device identical to the one used in the experiment.  
(b) A typical pulse trace from the output of the Agilent 81134A pulse generator, with pulse width $t_\mathrm{p}$.
(c) Diagram of energy levels of the system versus detuning $\varepsilon$. The (2,1) state \ket{0} anticrosses with the (1,2) states \ket{2} and \ket{3} with tunnel couplings $\Delta_1$ and $\Delta_2$. The two anticrossings are separated by an energy of $\delta E_\mathrm{R}$, which is the singlet-triplet energy splitting in the right dot.  Colors correspond to features in the data described in (e).
(d) Measurement of the transconductance through the QPC, which reflects changes in the charge occupation
of the double dot, as a function of pulse duration $t_\mathrm{p}$ and detuning of the pulse tip $\varepsilon_\mathrm{p}$.
(e)  Data from (d) in which different oscillation frequencies are highlighted in color.
The orange features at small detuning with frequency $\sim$5~GHz are 
charge qubit oscillations~\cite{Nakamura:1999p786,Petta:2004p1586,Dovzhenko:2011p161802,Petersson:2010p246804,Shi:2012preprint}
between the states \ket{0} and \ket{2}.
The pink features with frequency $\sim$9~GHz that occur at larger values of the detuning reflect  phase winding between states \ket{2} and \ket{3}.
(f) Results of the calculated quantum dynamics (see Methods for details)
of a system with $\delta E_\mathrm{R}=9.2~\mathrm{GHz}$ and tunnel couplings $\Delta_\mathrm{1}=2.62~\mathrm{GHz}$ and $\Delta_\mathrm{2}=3.5~\mathrm{GHz}$, including low-frequency noise in  detuning as in Ref.~\cite{Petersson:2010p246804}.
(g) Left: Bloch sphere of the projection of the wavefunction onto the \ket{2}, \ket{3} subspace with the trajectory of the state vector during the $\varepsilon_\mathrm{p}$-portion of the pulse mapped out for the case of $\varepsilon_\mathrm{p}=100~\mathrm{\mu eV}$. Right: the relative position of the pulse and the energy diagram for the data point labeled with pink pentagon ($\varepsilon_\mathrm{p}=100~\mathrm{\mu eV}$, $ t _\mathrm{p}=800~\mathrm{ps}$) in (f).
(h) Top: computed time evolution of the diagonal elements of the density matrix during the pulse for the data point labeled with pink pentagon in (f). The rising edge of the pulse increases the population of states \ket{2} and \ket{3} to 70\% and 23\% respectively.  Bottom: Time evolution of off-diagonal terms in the density matrix for the data point labeled with pink pentagon in (f). Relative phase oscillations between the two states during the $\varepsilon_\mathrm{p}$-portion of the pulse are clearly visible.
(i) The relative phase $\theta_{23}$ of states \ket{2} and \ket{3}, taken at the half point of the falling edge of the pulse, as a function of pulse width and the probability of measuring (1,2) charge occupation as a function of pulse width. The two curves are well correlated with each other, indicating the phase oscillation information during the pulse is mapped to charge probability by the falling edge of the pulse.
}
\label{fig1} 
\vspace{-8pt}
\end{figure*}

Fig.~\ref{fig1} demonstrates that oscillations both between states \ket{0} and \ket{2} as well as between states \ket{2} and \ket{3} can be
established and measured by application of the simple pulse sequence shown in Fig.~\ref{fig1}(b), which has been used in
previous work to investigate quantum oscillations between states of 
a charge qubit~\cite{Nakamura:1999p786,Petta:2004p1586,Dovzhenko:2011p161802,Petersson:2010p246804,Shi:2012preprint}.
The detuning voltage starts at a negative base value, where state \ket{0} is favored energetically (see Fig.~1c),
and then is pulsed to more positive detuning, close to the \ket{0}-\ket{2} and \ket{0}-\ket{3} anticrossings.
After a short time (typically of order 1-10 ns), the pulse ends and the detuning returns to its base value.
Fig.~\ref{fig1}(d) shows the resulting transconductance of the quantum point contact (QPC) indicated on Fig.~1(a),
which is sensitive
to changes in the time-averaged charge occupation of the dot.
If the electron is in one of the (1,2) states, \ket{2} or \ket{3}, at the end of the pulse, it remains in that state
until it decays inelastically back to a (2,1) state, which takes $\sim$18 ns~\cite{Shi:2012preprint}.
Thus, the average charge distribution in the dot reflects the occupation of the dot just after the end
of the pulse.
Two different types of
oscillations are observed and are highlighted in orange and pink in Fig.~\ref{fig1}(e): the former occur near the anticrossing between \ket{0} and \ket{2} and have a frequency that depends strongly on detuning.  The latter arise for $\varepsilon_\mathrm{p}$ more positive than the former location and have a frequency that is nearly independent of detuning.

\begin{figure*}[tb!]
\includegraphics[width=18cm]{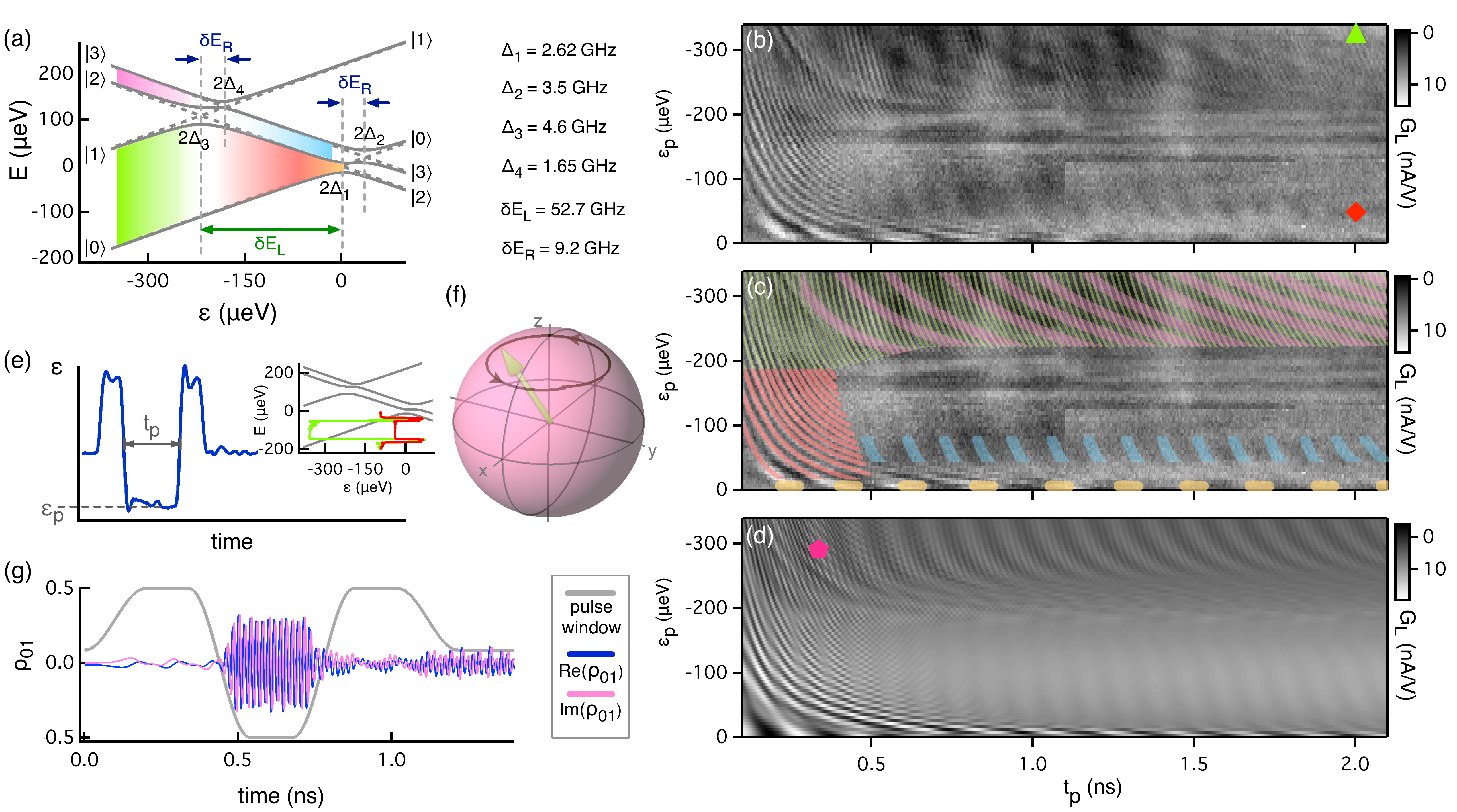}
\caption{\label{fig2} 
A simple multilevel pulse sequence induces quantum oscillations between different pairs of levels
at different values of the pulse detuning. 
(a) Diagram of relevant energy levels of the electrons in the double quantum dot.  The oscillations highlighted in the measurements shown in panel (c) correspond to the transitions between levels denoted by the appropriately colored regions in this diagram. The parameters used in the simulation are also listed. 
(b) Measured transconductance through the charge sensing QPC, which reflects changes in the time-averaged charge state of the double quantum dot as a function of the pulse detuning $\varepsilon_\mathrm{p}$ and of the pulse duration $t_\mathrm{p}$, with all other parameters held fixed, in the presence of the pulse sequence in (e).  Coherent oscillations between different pairs of charge states are reflected in the oscillation of the time-averaged charge occupation in the dot as a function of  $t_\mathrm{p}$ at different values of the detuning.
(c) Same as (b) with the different frequencies highlighted by differently colored lines. The oscillations highlighted here correspond to the transitions between levels denoted by the corresponding colored regions in (a).
(d) Results of numerical simulation of the system with parameters listed in (a), in the presence the pulse sequence in (e). The pulse rise time used is $118~\mathrm{ps}$. 
(e) A typical pulse trace for this experiment. Inset: The relative position of the pulses with respect to the energy levels for the data point labeled by the green triangle and red diamond in (b). 
(f) Projection onto the Bloch sphere for the states \ket{0} and \ket{1}, with the trajectory of the state vector during the $\varepsilon_\mathrm{p}$-portion of the pulse mapped out at $\varepsilon_\mathrm{p} = -291 ~\mathrm{\mu eV}$.
(g) Computed time evolution of the coupling term between states $|0\rangle$ and $|1\rangle$ for the data point labeled with pink pentagon in (d). Relative phase oscillations between the two states during the $\varepsilon_\mathrm{p}$-portion of the pulse are clearly visible. 
}
\vspace{-8pt}
\end{figure*}

To gain insight into the two different oscillation frequencies shown in Fig.~\ref{fig1}(d-e),
we perform numerical simulations (see Methods for details)
of the dynamics of a system with
the energy spectrum shown in Fig.~\ref{fig2}(a), with low-frequency detuning noise incorporated as
in Ref.~\cite{Petersson:2010p246804}.
Fig.~\ref{fig1}(f) shows the result of the simulation, which is in good agreement with the data.
When $\varepsilon_\mathrm{p}\approx 0$, the oscillations (highlighted
with orange in Fig.~\ref{fig1}(e))
are between the states \ket{0} and \ket{2}, the ground states of the (2,1) and (1,2) charge occupations.
The ``sideways-v'', criss-cross pattern of the oscillations in this regime is characteristic of lock-in measurement of 
charge qubits~\cite{Petta:2004p1586,Dovzhenko:2011p161802,Petersson:2010p246804,Shi:2012preprint};
it arises because the oscillation frequency depends strongly on $\varepsilon_\mathrm{p}$, with a minimum
frequency of $2\Delta_1$ at $\varepsilon_\mathrm{p}=0$.
At larger values of $\varepsilon_\mathrm{p}$, oscillations at a different frequency appear (highlighted by the nearly parallel pink lines near the
bottom of Fig.~\ref{fig1}(e)).
These oscillations have a different period ($\sim 100$~ps) that depends only weakly on $\varepsilon_\mathrm{p}$; they are well-described by the simulation of Fig.~\ref{fig1}(f), 
and their frequency is set by the energy difference between the states \ket{2} and \ket{3}.
As is clear from the full time evolution of each relevant state, which is plotted in Fig.~\ref{fig1}(h), at this detuning the rising edge of the pulse transfers the large majority of the weight in the wavefunction into states \ket{2} and \ket{3}, leaving very little occupation of \ket{0}.  An approximate quantum wavefunction during the $\varepsilon_\mathrm{p}$-portion of the pulse is thus given by
\begin{equation}
\label{eq:zrotation}
\ket{\psi(t)} \approx e^{i\phi(t)}
\left ( 
a\ket{2}
+be^{i\delta E_\mathrm{R}t/\hbar} 
\ket{3}
\right )~,
\end{equation}
with $\phi(t)$ a global phase that does not affect measurable quantities.
While the charge sensing measurement does not distinguish between states \ket{2} and \ket{3},
the oscillations are visible in this experiment because the two contributions
interfere when the pulse ends and the detuning passes back through the two anticrossings shown in Fig.~1(c), between
\ket{0} and \ket{2} as well as \ket{0} and \ket{3}, so that the occupation of the (1,2) charge state after the pulse has ended
oscillates with angular frequency $\delta E_\mathrm{R}/\hbar$.
This relationship between the phase difference between the amplitudes in \ket{2} and \ket{3} and
the probability of occupying the (1,2) charge state $P_{(12)}$ is illustrated in Fig.~\ref{fig1}(i).
The physical mechanism giving rise to the ability to measure oscillations between two states with the
same charge distribution via a time-averaged charge measurement
is closely related to Landau-St\"uckelberg-Zener
oscillations~\cite{Shevchenko:2010p1,Stehlik:2012p121303,Cao:2013p1401},
so we will refer to these oscillations as LSZ oscillations.

The figure of merit (the ratio of the coherence time $T_2^*$, extracted from the oscillation decay at times longer than those shown in Fig.~1(d), to the oscillation period) of the LSZ oscillations 
between \ket{2} and \ket{3}
is much larger than that of the charge qubit oscillations between \ket{0} and \ket{2}, for two reasons.  
First,  the frequency of the LSZ oscillations is not limited by
the pulse rise time; the oscillation frequency is determined by the energy difference $\delta E_\mathrm{R}$ between state \ket{2} and \ket{3};
whether a particular pulse rise time results in a state in which \ket{2} and \ket{3} both have substantial occupation is determined
by the value of the tunnel couplings $\Delta_1$ and $\Delta_2$ .
Second, the energy difference $\delta E_\mathrm{R}$ depends only weakly on detuning~\cite{Shi:2011p233108}, so
the LSZ oscillations are
 less susceptible to the dominant source of decoherence~\cite{Dial:2013p146804,Petersson:2010p246804,Dovzhenko:2011p161802}, fluctuations in the detuning, than are
oscillations between energy levels with different dependences on the detuning, such as the
levels \ket{0} and \ket{2} used for a standard charge qubit.
 
We now show that more complex pulse sequences can establish oscillations between different
pairs of states in the system, including between states that are both excited states of the system
at all values of the detuning accessed during the sequence.
Fig.~\ref{fig2}(b) shows the measured transconductance of the QPC charge sensor during application of the voltage pulse shown in Fig.~\ref{fig2}(e).  Five distinct oscillation patterns can be identified, as shown with the color overlay in Fig.~2(c), and with corresponding colors in Fig.~2(a).  The numerical simulation of Fig.~2(d), which uses the same parameters as the simulation in Fig.~1(f), shows that the oscillation frequencies in the data correspond to energy level differences between specific pairs of quantum states.  That mapping is shown in Fig.~2(a), and the validity of our model is demonstrated by the accuracy of Fig.~2(d).  The color overlays in Fig.~2(a) and 2(c) show that, even with a single, relatively simple pulse pattern, quantum superpositions and oscillations can be observed between nearly all possible pairs of states.  The relative weight of each of these oscillations, which reflects the relative weight of the wavefunction in each rung in the ladder of energy eigenstates, is determined by the pulse rise time, the pulse detuning, and the tunnel couplings that determine the size of the anticrossings between the three-electron states.

{\em Relationship to the quantum dot hybrid qubit.}
One important reason for manipulating quantum states is to perform quantum information processing.
For this application, one needs to create quantum gates, which are unitary transformations.
If one defines a qubit as two states $\ket{\tilde{0}}$ and $\ket{\tilde{1}}$, then to qualify as a gate, a process
that transforms $\ket{\tilde{0}} \rightarrow a\ket{\tilde{0}}+b\ket{\tilde{1}}$ must transform either $\ket{\tilde{1}} \rightarrow -b^*\ket{\tilde{0}}+a^*\ket{\tilde{1}}$
or $\ket{\tilde{1}} \rightarrow b^*\ket{\tilde{0}}-a^*\ket{\tilde{1}}$.
Ref.~\cite{Koh:2012p250503} presents pulse sequences that, when applied to a double quantum dot with three electrons,
yield any prescribed rotation on the Bloch sphere of a qubit with basis states $\ket{\tilde{0}}=\ket{0}$, $\ket{\tilde{1}}=\ket{1}$, where \ket{0} and \ket{1} are two of the states we study here.
A $\pi$ rotation that sends $\ket{0}\rightarrow\ket{1}$ 
can be implemented by performing successive $\pi$ rotations at the two successive
anticrossings marked $\Delta_1$ and $\Delta_3$ in
Fig.~2(a).

Here we can understand the oscillations highlighted green in Fig.~2(c) in the language of the hybrid qubit.  The first 340~ps pulse in Fig.~2(e) rotates significant weight of the wavefunction from \ket{0} into state \ket{2}, which would be called an \emph{auxilliary state} in a (2,1) hybrid qubit.  The second, variable section of the pulse pushes the double dot to deep negative detuning, with very different effects on the fraction of the wavefunction in states \ket{0} and \ket{2}.  State \ket{0} simply slides to lower energy in the detuning plot shown in Fig.~2(a). State \ket{2}, in contrast, moves to higher energy, where it anticrosses with  \ket{1}.  This anticrossing, governed by tunnel coupling $\Delta_3$, is large enough that the pulse is largely adiabatic and therefore the majority of the weight in \ket{2} follows the lower branch to state \ket{1}, whose dependence of energy on detuning is nearly the same as \ket{0}, setting up a superposition whose phase difference is relatively immune to noise in detuning.  The second 340~ps pulse in Fig.~2(e) reverses this process and drives a second rotation at the $\Delta_1$ anticrossing, enabling observation of interference as a function of the evolved phase difference between states \ket{0} and \ket{1}.

The figure of merit for the resulting oscillations is over 100, an extremely high value for the present state of semiconductor qubits.  Thus, the oscillations shown in green reflect a controlled phase evolution between states \ket{0} and \ket{1}, a $\hat{z}$-rotation for the hybrid qubit, demonstrating a key ingredient in constructing a pulse-gated quantum dot hybrid qubit~\cite{Koh:2012p250503}.

We have demonstrated that high-speed voltage pulses can be used
to control coherent quantum oscillations
between different pairs of states in a semiconducting double quantum dot with three electrons.
By implementing appropriate combinations of voltage pulses, oscillations between different pairs
of levels as well as sequential operations can be achieved.
Transitions between some pairs of levels can be induced that have over a hundred oscillations
within a coherence time.  All of the observed rotations and oscillations have frequencies in excess of 5~GHz.
These results provide strong evidence that coherent, fast oscillations can be initiated and controlled
between multiple levels of three electrons in a double quantum dot.

\textbf{Methods.}~~\emph{Experiment:}
The experiments are performed on a double quantum dot fabricated in a Si/SiGe
heterostructure, as described in Ref.~\cite{Thalakulam:2010p183104,Simmons:2011p156804,Shi:2012preprint};
a scanning electron microscope image of an identical device  is shown in Fig.~\ref{fig1}(a).
By adjusting the gate voltages appropriately,
we tune the dot occupation so that the valence charge occupation~\footnote{Either or both dots 
may contain a closed shell beneath the valence electrons; 
if present, such shells do not appear to play a role in the work we report.} of the double dot is 
(2,1) or (1,2), where the first (second) number is the charge occupation in the left (right) dot, as confirmed
by magnetospectroscopy measurements~\cite{Shi:2011p233108}. 

\emph{Theory:} Numerical simulations of the experiment were performed based on the energy level diagram in Fig.~\ref{fig2}(a)
using the a pulse rise time of 118 ps.  
We model the dynamical evolution of the density matrix $\rho$ of the system as a function of detuning $\varepsilon$ and pulse duration $t_\mathrm{p}$ using a master equation~\cite{NielsenBook}:
\begin{equation}
\dot{\rho} = -\frac{i}{\hbar} [ H , \rho ]
\end{equation}
The Hamiltonian, written in the basis of position eigenstates, is

\begin{equation}
H = \left( 
\begin{array} {cccc}
\varepsilon/2 & 0 &\Delta_1& -\Delta_2 \\
0&\varepsilon/2+\delta E_\mathrm{L}&-\Delta_3 &\Delta_4 \\
\Delta_1&-\Delta_3&-\varepsilon/2  &0 \\
-\Delta_2&\Delta_4 & 0 &-\varepsilon/2+\delta E_\mathrm{R} \\
\end{array} \right).
\end{equation}
The (1,2) charge occupation probability is extracted at the end of the pulse and is averaged over 2~ns in the measurement stage of the pulse. Using this number as an initial value, $P_\mathrm{(1,2)}$ is then allowed to relax exponentially to the ground state (2,1) occupation with a relaxation time $T_1$, during the measurement phase. Finally, the simulated charge occupation is determined by averaging the charge state for the entire 33~ns pulse period.
Low-frequency fluctuations in the detuning $\varepsilon$ are incorporated following Ref.~\onlinecite{Petersson:2010p246804}, by performing a convolution of the results at each $\varepsilon$ with a Gaussian in $\varepsilon$ of width  $\sigma_\varepsilon = 5~\mu$eV.  The best fit to the data is found with a charge $T_1=18$~ns.


\bibliography{siliconqcsnc}

\emph{Acknowledgements.}~~~
This work was supported in part by ARO (W911NF-08-1-0482, W911NF-12-0607), by NSF (PHY-1104660), and by the United States Department of Defense.
The views and conclusions contained in this document are those of the authors and should not be interpreted as representing the official policies, either expressly or implied, of the US Government.  
Development and maintenance of the growth facilities used for fabricating samples is supported by DOE (DE-FG02-03ER46028).
This research utilized NSF-supported shared facilities at the University of Wisconsin-Madison.

\emph{Author contributions.}~~~
C.B.S., D.E.S, M.G.L., M.A.E. designed and fabricated the sample, Z.S., D.R.W., J.R.P., R.T.M., X.W., S.N.C., M.A.E. performed the experiment and analyzed the data, Z.S., T.S.K., J.K.G., M.F., S.N.C. performed the calculations, Z.S., M.F., S.N.C., M.A.E. wrote the paper.
 
\emph{Competing Interests.}~~~The authors declare that they have no
competing financial interests.

\emph{Correspondence}~~~Correspondence should be addressed to M.A.E.~(email: maeriksson@wisc.edu).


%

\end{document}